\documentclass[11pt]{article}
\usepackage{amsmath,amssymb,amsthm,mathtools}
\usepackage{array}
\usepackage{graphicx}
\usepackage{booktabs}
\usepackage[round,authoryear]{natbib}
\usepackage{float}
\usepackage{caption}
\usepackage{needspace}
\usepackage{placeins}
\usepackage{tabularx}
\usepackage{xcolor}
\usepackage[colorlinks,citecolor=blue,linkcolor=blue,urlcolor=blue]{hyperref}

\newcolumntype{Y}{>{\raggedright\arraybackslash}X}

\DeclareMathOperator{\Var}{Var}

\newcommand{\E}{\mathbb{E}}
\newcommand{\Pp}{\mathbb{P}}
\newcommand{\R}{\mathbb{R}}
\newcommand{\one}{\mathbf{1}}
\newcommand{\argmin}{\operatorname*{argmin}}
\newcommand{\dd}{\operatorname{d}}
\newcommand{\psitau}{\psi_\tau}
\newcommand{\norm}[1]{\lVert #1\rVert}
\newcommand{\Gcal}{\mathcal G}

\newtheorem{theorem}{Theorem}
\newtheorem{corollary}{Corollary}
\newtheorem{proposition}{Proposition}
\theoremstyle{remark}
\newtheorem{remark}{Remark}
\numberwithin{equation}{section}

\hypersetup{pdftitle={Joint Estimation and Grouping of Quantile Association Curves across Multiple Responses},pdfauthor={Tianying Wang, Yanyuan Ma, Ying Wei}}

\begin{document}

\title{Joint Estimation and Grouping of Quantile Association Curves across Multiple Responses}

\author{Tianying Wang\thanks{Department of Statistics, Colorado State University, Fort Collins, CO 80523, USA. \href{mailto:Tianying.Wang@colostate.edu}{Tianying.Wang@colostate.edu}}
\quad and \quad Yanyuan Ma\thanks{Department of Statistics, Pennsylvania State University, University Park, PA 16802, USA.}
\quad and \quad Ying Wei\thanks{Department of Biostatistics, Columbia University, New York, NY 10032, USA.}}

\date{}

\maketitle
\thispagestyle{empty}
\baselineskip=20pt

\begin{abstract}
Large biobanks provide an opportunity to study how associations with
related traits vary across age and across their distributions. We propose a
quantile regression procedure that jointly groups responses and estimates
their shared varying-coefficient curves on conditional percentile scales.
An information criterion selects both the number of groups and their
membership from grouped B-spline fits.
We establish consistent group recovery, curve convergence rates, and feasible
pointwise inference after group selection as the spline dimension grows.
The analysis distinguishes true curve differences from improvements due to
separate spline approximations of a shared curve. Errors in the estimated
percentile scales partly cancel in comparisons of pooled and split fits of a
shared curve, but can still contribute to the asymptotic variance of the curve
estimators. In 127{,}040 UK Biobank women aged 40 to 70, we grouped
nine metabolic traits by their associations with a type 2 diabetes polygenic
risk score and found seven groups at the median, compared with three and
four at the lower and upper quantiles. The method identifies shared tail
associations among traits with distinct median associations, allowing their
information to be pooled where the curves agree.

\end{abstract}

\pagestyle{plain}

\section{Introduction}
\label{sec:introduction}

Many studies measure multiple related responses in individuals of different ages.
Their associations with a common predictor may vary with age and across the
response distribution \citep{bryois2017time,wang2025regenieqrs}.
One common objective is to determine which responses share a common association
pattern and which exhibit distinct patterns.

Genetic association studies provide a concrete example of this problem. Genetic
variants can influence multiple phenotypes through shared biological pathways,
yet the resulting associations may emerge at different ages or change in
different ways over time. These patterns have a biological basis: cross-trait
genome-wide analyses have identified genetic loci associated with multiple
traits \citep{watanabe2019pleiotropy}, while the regulation of some genes
changes over time \citep{bryois2017time} and circulating proteins exhibit
nonlinear age-related profiles \citep{lehallier2019proteome}. Systematically
characterizing such cross-trait and age-dependent associations remains
challenging, in part because statistical methods for jointly identifying shared
association patterns across responses and over time on a common percentile scale
are limited.

Developing such a framework requires addressing several features. First, the
responses are often measured on different scales, making direct comparison of
their association coefficients difficult \citep{thurston2009bayesian}. Second,
the association between the predictor and each response often varies with age or
time, so a single constant-effect model may miss important temporal structure
\citep{hastie1993varying,wang2009quantile}. In addition, these associations may
differ across the conditional response distribution: effects near the center may
not represent effects in the lower or upper tails
\citep{koenker2005quantile,wang2025regenieqrs}.

These features motivate a joint varying-coefficient quantile-regression
framework for discovering shared association patterns over time across responses.
We first standardize the responses using age-specific percentile ranks, placing
responses measured on different scales on a common relative scale. Age-specific
percentile positions have a familiar interpretation in reference-centile methods
\citep{cole1992centile}. At each quantile level, we then model the association
between the predictor and each rank-standardized response as a smooth function of
age, while allowing the coefficient function to be shared by an unknown group of
responses. We estimate the number of groups, their memberships, and the shared
association curves (approximated by B-splines) jointly by minimizing a penalized
grouped quantile-loss criterion. Each resulting curve describes how the
predictor is associated with a response's relative position within the reference
population at a given age. The estimated number and composition of the groups
reveal how many distinct age-dependent association patterns are present and
which responses exhibit coordinated patterns at each quantile level.

Related work on multiple responses includes mean-based response grouping
\citep{zhou2025heterogeneous}, response clustering based on quantile-indexed
coefficient curves \citep{sottile2019clusters}, estimation of shared components
across response-specific curves \citep{jung2026reduced}, and covariate selection
\citep{park2023regional}.
Other work on learning shared associations has primarily studied heterogeneous
association structures \textit{across subjects} or \textit{regions of the
covariate space}, such as
\citet{liu2020identification,leng2021multi,zeileis2008model,su2019sieve,zhang2023nonparametric},
\citet{gu2018panel,yang2024functional}, among many.
Our problem is distinct: we partition multiple responses according to their
shared predictor-association curves over age or time at a fixed quantile level,
after placing responses measured on different scales on a common percentile
scale.
Although our work was motivated by genetic association studies,  the same
statistical problem arises more broadly whenever associations between a predictor and responses
vary with a continuous variable
such as age, time, or location. The goal is
to identify groups of responses that share an association curve across a
specified part of the response distribution, while allowing the grouping
structure to differ across quantile levels.

We develop a methodological framework for learning shared quantile associations
across responses and over time, with theoretical guarantees for data-driven
response grouping and inference. We prove consistency of the estimated number of
groups and their memberships, derive convergence rates for the shared B-spline
coefficient curves, and establish valid pointwise inference as the spline
dimension increases with the sample size. These results remain valid when both
the percentile ranks and the response
groups are estimated from the data.

The remainder of the paper is organized as follows. Section~\ref{sec:method}
presents the proposed method, and Section~\ref{sec:theory} establishes its
theoretical properties. Sections~\ref{sec:simulation_generated_rank}
and~\ref{sec:ukb-analysis} report simulation studies and an application to UK
Biobank data, respectively. Section~\ref{sec:discussion} concludes with a
discussion. Proofs and additional computational and numerical details are
provided in the Supplementary Material.

\section{Joint Estimation and Selection of Shared Curves}
\label{sec:method}

At each fixed quantile level, we jointly estimate the response partition and
the coefficient curves shared within its groups. Conditional percentiles
define the scale on which curves are compared, and a penalized grouped
quantile-regression criterion selects the partition. Pointwise inference for
the selected curves accounts for estimation of these percentiles.

\subsection{Conditional-Rank Target}
\label{subsec:setup}

We place responses on a common scale by expressing each
observed value as its percentile position in a conditional response
distribution. In the age
example, this is its position among individuals of the same age. We then model
how these percentile positions are associated with the predictor.
We observe \(n\) independent vectors
\(O_i=(O_{i1},\ldots,O_{iK})\), each containing \(K\) responses and their
covariates, with \(K\) fixed. The responses within each vector may be dependent.

For the \(k\)th response, write
\(O_{ik}=(\widetilde Y_{ik},X_{ik},C_{ik},T_{ik})\), where
\(\widetilde Y_{ik}\in\R\), \(X_{ik}\in\R\), \(C_{ik}\in\R^q\), and
\(T_{ik}\in[0,1]\), with fixed \(q\). The conditional distribution used
for rank standardization is
\(F_{0k}(y\mid t)=\Pp(\widetilde Y_{ik}\le y\mid T_{ik}=t)\), and the
oracle conditional rank is \(R_{ik}=F_{0k}(\widetilde Y_{ik}\mid T_{ik})\).
When \(F_{0k}(\cdot\mid t)\) is continuous, \(R_{ik}\mid T_{ik}\)
is uniform on \((0,1)\). The
conditional distribution of \(R_{ik}\) given \((X_{ik},C_{ik},T_{ik})\) need
not be uniform. The model below is therefore a regression model for the
conditional rank distribution given the covariates.

We fix a quantile level \(\tau\in(0,1)\) and use the exact-sharing rank-scale
model
\[
Q_{R_k}(\tau\mid X_{ik},C_{ik},T_{ik})
=
X_{ik}\alpha_{g_0(k)}^0(T_{ik})+C_{ik}^{\top}\xi_k^0 .
\]
Here \(g_0(k)\in\{1,\ldots,M_0\}\) is the unknown group membership,
\(\alpha_m^0\) is the common coefficient function for group \(m\), and
\(\xi_k^0\) is a response-specific finite-dimensional coefficient vector. Let
\(G_m^0=\{k:g_0(k)=m\}\) and
\(\Gcal_0=\{G_1^0,\ldots,G_{M_0}^0\}\). Exact sharing means equality of the
rank-scale coefficient functions within a group. No equality is imposed on the
raw response scale. Our primary targets are the partition
\(\Gcal_0\) and its shared coefficient functions
\(\alpha_1^0,\ldots,\alpha_{M_0}^0\).

The conditional distribution \(F_{0k}\) is unknown. We estimate it separately
for each response and use cross-fitting.
Let \(I_1,\ldots,I_L\) be fixed, equal-size folds of observation
indices, chosen independently of the data, with \(L\) fixed. Let
\(\ell(i)\) denote the fold containing observation \(i\), and let
\(\widehat F_k^{(-\ell(i))}\) be a proper
conditional distribution estimator constructed without fold \(\ell(i)\). The
generated rank used in the second stage is
\(\widehat R_{ik}=\widehat F_k^{(-\ell(i))}(\widetilde Y_{ik}\mid T_{ik})\).

The theory uses an integrated-kernel conditional distribution estimator. With
fixed \(L\) folds, one concrete choice is
\[
\widehat F_k^{(-\ell)}(y\mid t)
=
\frac{\sum_{j:\ell(j)\ne \ell}K_h(T_{jk}-t)\mathcal K_{b_y}(y-\widetilde Y_{jk})}
{\sum_{j:\ell(j)\ne \ell}K_h(T_{jk}-t)},
\qquad
\mathcal K_{b_y}(a)=\int_{-\infty}^{a/b_y}K_y(s)\dd s .
\]
If the denominator is zero, we average
\(\mathcal K_{b_y}(y-\widetilde Y_{jk})\) with equal weights over observations
\(j\) outside fold \(\ell\), that is, those with \(\ell(j)\ne\ell\).
Here \(K_h(u)=h^{-1}K_T(u/h)\), and \(h\) and \(b_y\) are the smoothing
bandwidths in \(T\) and \(y\), respectively. The nonnegative
distribution-function kernel gives an estimator that is monotone in \(y\) and
takes values in \([0,1]\). Because the estimator uses observations outside
\(I_\ell\), it is fixed conditional on those observations when we analyze ranks
within \(I_\ell\) \citep{chernozhukov2018double}. Other first-stage estimators
require verification of the score and refinement loss expansions in Supplement
S.2 and S.5, in addition to the first-stage rates in Section~\ref{sec:theory}.

\subsection{Grouped Estimation and Information Criterion}
\label{subsec:generated-estimation}

We estimate a shared curve by pooling the responses assigned to
its group. For a candidate partition \(\Gcal=\{G_1,\ldots,G_M\}\), let
\(B(t)\in\R^{J_n}\) be a B-spline basis. If \(k\in G_m\), we approximate
\(\alpha_m(t)\) by \(B(t)^\top\theta_m\) and define
\(\vartheta_{\Gcal}=(\theta_1^\top,\ldots,\theta_M^\top,\xi_1^\top,\ldots,\xi_K^\top)^\top\).
The grouped design vector \(D_{ik,\Gcal}\) is defined by the identity
\(D_{ik,\Gcal}^{\top}\vartheta_{\Gcal}
=X_{ik}B(T_{ik})^\top\theta_m+C_{ik}^{\top}\xi_k\).
For a fixed partition, the estimator based on the generated ranks is
\begin{equation}
\label{eq:fixed-partition-estimator}
\widehat\vartheta_{\Gcal}
\in
\argmin_{\vartheta_{\Gcal}}
\frac{1}{n}\sum_{i=1}^n\sum_{k=1}^K
\rho_\tau\{\widehat R_{ik}-D_{ik,\Gcal}^{\top}\vartheta_{\Gcal}\},
\end{equation}
where \(\rho_\tau(u)=u\{\tau-\one(u<0)\}\) and
\(\psitau(u)=\tau-\one(u<0)\). The oracle spline pseudo-target
\(\vartheta_{\Gcal}^0\) is the population minimizer of the same criterion with
\(\widehat R_{ik}\) replaced by \(R_{ik}\).

For \(Z\in\{R,\widehat R\}\), define
\[
Q_n^Z(\Gcal,\vartheta_{\Gcal})
=
\frac{1}{n}\sum_{i=1}^n\sum_{k=1}^K
\rho_\tau\{Z_{ik}-D_{ik,\Gcal}^{\top}\vartheta_{\Gcal}\},
\qquad
Q^R(\Gcal,\vartheta_{\Gcal})
=
\E\sum_{k=1}^K
\rho_\tau\{R_{ik}-D_{ik,\Gcal}^{\top}\vartheta_{\Gcal}\}.
\]

We select the partition by penalizing the minimized loss for the number of
estimated coefficients. Let
\([K]=\{1,\ldots,K\}\), let \(\mathfrak P_M([K])\) denote the set of all
unlabeled partitions of \([K]\) into \(M\) nonempty blocks, and let
\(\mathcal M\subseteq\{1,\ldots,K\}\) be a fixed nonempty set of candidate
group numbers. These choices define the finite candidate family
\(\mathfrak G=\bigcup_{M\in\mathcal M}\mathfrak P_M([K])\). Let
\(L_n(\Gcal)=\inf_{\vartheta_{\Gcal}}
Q_n^{\widehat R}(\Gcal,\vartheta_{\Gcal})\). With
\(d(\Gcal)=|\Gcal|J_n+Kq\), let \(a_n>0\) be the complexity-penalty
multiplier, and define
\begin{equation}
\label{eq:ic-selector}
IC_n(\Gcal)=L_n(\Gcal)+a_nd(\Gcal),
\qquad
\widehat\Gcal\in\argmin_{\Gcal\in\mathfrak G}IC_n(\Gcal),
\qquad
\widehat M=|\widehat\Gcal|.
\end{equation}

\subsection{Exact Partition Search}
\label{subsec:exact-cached-selection}

We minimize the information criterion by reusing the fitted loss
for each possible group. Because the loss for a partition is the sum of its
group losses, we need to fit each group only once. For every nonempty subset
\(A\subseteq[K]\), define its profiled grouped loss by
\begin{equation}
\label{eq:subset-profiled-loss}
\widehat L_\tau(A)
=
\min_{\theta_A,\{\xi_k:k\in A\}}
\frac{1}{n}\sum_{i=1}^n\sum_{k\in A}
\rho_\tau\{\widehat R_{ik}
-X_{ik}B(T_{ik})^\top\theta_A-C_{ik}^\top\xi_k\},
\end{equation}
where the curve coefficient \(\theta_A\) is shared by the responses in
\(A\), whereas the nuisance coefficient \(\xi_k\) remains response
specific. Because the check-loss criterion is convex,
each subset fit is a convex quantile-regression problem
\citep{koenker1978regression,koenker2005quantile}.

For \(\mathcal G=\{G_1,\ldots,G_M\}\in\mathfrak P_M([K])\), the profiled loss is
\begin{equation}
\label{eq:partition-additivity}
L_n(\mathcal G)
=\sum_{m=1}^M\widehat L_\tau(G_m).
\end{equation}
Minimization separates across the disjoint group-specific parameter blocks,
so the fit for a subset \(A\) can be reused in every partition containing it.
For example, with three responses and all group numbers allowed,
the \(2^3-1=7\) group fits cover the three single responses, the three pairs,
and the group of all three. We compare the five possible partitions by
adding their group losses and penalties, using
\(\widehat L_\tau(\{1,2\})+\widehat L_\tau(\{3\})\) for the loss of
the partition \(\{\{1,2\},\{3\}\}\).

The exact search has two stages. First, we compute and store
\(\widehat L_\tau(A)\) and its minimizer for every nonempty subset
\(A\subseteq[K]\) that can occur in a candidate partition. When all group
numbers are allowed, this requires \(2^K-1\) fits. Second, we combine the
stored losses using the dynamic-programming principle of optimality
\citep{bellman1957dynamic}.
For a nonempty \(S\subseteq[K]\) and an integer \(m\leq |S|\), let
\(V_\tau(S,m)\) be the smallest total stored loss among all partitions of \(S\)
into \(m\) blocks. The base case is
\(V_\tau(S,1)=\widehat L_\tau(S)\). For \(m\geq2\), we use the recursion
\begin{equation}
\label{eq:partition-dp}
V_\tau(S,m)
=
\min_{\substack{A\subset S:\min(S)\in A,\\
                 |S\setminus A|\geq m-1}}
\left\{
\widehat L_\tau(A)+V_\tau(S\setminus A,m-1)
\right\}.
\end{equation}

The restriction \(\min(S)\in A\) avoids repeated block orderings. We recover the
partition and its coefficient estimate from the stored minimizers, using the
deterministic labeling and tie rule in Section~S.7. The following proposition
shows that the recursion finds the smallest loss for each candidate group number
when every subset problem is solved exactly.

\begin{proposition}[Finite-sample global optimality]
\label{prop:exact-cached-global}
Suppose every subset problem in \eqref{eq:subset-profiled-loss} is solved to
global optimality. For each $M\in\mathcal M$, recursion
\eqref{eq:partition-dp} returns a partition
$\widehat{\mathcal G}_{\tau,M}$ satisfying
\[
L_n(\widehat{\mathcal G}_{\tau,M})
=
\min_{\mathcal G\in\mathfrak P_M([K])}
L_n(\mathcal G).
\]
Consequently, combining the recursion with the information criterion returns a
global minimizer over the candidate space
\(\bigcup_{M\in\mathcal M}\mathfrak P_M([K])\).
\end{proposition}

The proposition connects the computed partition to the criterion minimizer
studied in Section~\ref{sec:theory}. The search covers the full candidate family
without requiring an initial grouping. In computation, each subset loss is
evaluated to the tolerance of the
quantile-regression solver. Section~S.7 gives the proof, algorithm, and
computational complexity.

\subsection{Effects of Rank Estimation on Scores and Losses}
\label{subsec:corrected-inference}

We now examine how estimating the percentile scales affects curve
uncertainty and the loss comparisons used to select a partition. For
inference, we account for two contributions from each independent
observation \(i\), which contains \(K\) responses. In addition
to its own quantile scores, observation \(i\) enters the training samples
used to rank observations in other folds. We combine these contributions in
the influence expansion. Let
\(S_{ik,\Gcal}=D_{ik,\Gcal}\psitau
\{R_{ik}-D_{ik,\Gcal}^{\top}\vartheta_{\Gcal}^0\}\).
Let \(W_{ik}=(X_{ik},C_{ik},T_{ik})\), and let
\(b_{ik}(v)\) be the conditional density of \(R_{ik}\) at \(v\),
given \(W_{ik}\). Section~S.1 expresses this density as a ratio of
raw-response densities. The density itself is common to all candidate
partitions. For each candidate, we evaluate it at the population spline fitted
value
\(v_{ik,\Gcal}=D_{ik,\Gcal}^{\top}\vartheta_{\Gcal}^0\). Write
\(\pi_{k,\Gcal}(v\mid t)\) for the conditional density of this fitted value at
\(v\), given \(T_{ik}=t\). We use the representer \(\lambda_{k,\Gcal}\) to
describe how
errors in the estimated reference CDF change the expected quantile score to
first order. The correction \(\Delta_{ik,\Gcal}\) accounts for observation
\(i\) when it is used to estimate ranks for observations in other folds.
Section~S.1 specifies a fixed rank interval inside \((0,1)\) containing
the candidate fitted values.
We set \(\lambda_{k,\Gcal}\) to zero outside the interior working interval
and define
\[
\begin{aligned}
\lambda_{k,\Gcal}(v,t)
&=
\pi_{k,\Gcal}(v\mid t)
\E\{D_{ik,\Gcal}b_{ik}(v)\mid
D_{ik,\Gcal}^{\top}\vartheta_{\Gcal}^0=v,T_{ik}=t\},
\\
\Delta_{ik,\Gcal}
&=
\int_0^1
\lambda_{k,\Gcal}(v,T_{ik})
\{\one(R_{ik}\le v)-v\}\dd v .
\end{aligned}
\]

We evaluate the conditional rank density at the population spline fitted value
to obtain \(b_{ik,\Gcal}=b_{ik}(v_{ik,\Gcal})\).
Equivalently, \(b_{ik,\Gcal}\) is the conditional
density of the residual
\(R_{ik}-D_{ik,\Gcal}^{\top}\vartheta_{\Gcal}^0\) at zero, given
\(W_{ik}\). The information
matrix is
\(A_{\Gcal}=\sum_{k=1}^K
\E\{b_{ik,\Gcal}D_{ik,\Gcal}D_{ik,\Gcal}^{\top}\}\).
Combining the oracle score with the rank-estimation correction gives the
contribution from observation \(i\)
\(\Phi_{i,\Gcal}=\sum_{k=1}^K
\{S_{ik,\Gcal}+\Delta_{ik,\Gcal}\}\),
with covariance
\(\Omega_{\Gcal}=\Var(\Phi_{i,\Gcal})\).
Together, \(A_{\Gcal}\) and \(\Omega_{\Gcal}\) determine the corrected
sandwich covariance for the coefficient estimator. We account for rank
estimation through
\(\Delta_{ik,\Gcal}\), including its covariance with the oracle score.
We form the variance after summing the contributions from all \(K\)
responses in observation \(i\), so the covariance also includes their
dependence.

For partition selection, we compare candidate losses using the
same estimated ranks, so common first-stage errors partly cancel for pooled and
split fits of a true response group. Section~\ref{subsec:selection-refinement}
shows how this cancellation
and the approximation gain from splitting a true group determine the penalty
requirements; Section~S.5 gives the loss comparison.

\subsection{Pointwise Multiplier Bootstrap}
\label{subsec:feasible-inference}

We construct a pointwise interval for the spline curve value of a
specified response at a fixed \(t\), or more generally for a scalar
coefficient contrast. For every candidate partition \(\Gcal\), we fix a
contrast vector \(c_{n,\Gcal}\) in its coefficient space. To evaluate a
curve, this vector has \(B(t)\) in the block of the group containing that
response and zeros elsewhere. We write
\(c_n=c_{n,\Gcal_0}\) and
\(\widehat c_n=c_{n,\widehat\Gcal}\). This convention makes the selected
statistic well-defined when candidate dimensions differ. The target is
\(c_n^\top\vartheta_{\Gcal_0}^0\). Section~\ref{sec:theory} states the
additional
bias conditions for replacing a spline pseudo-target by the corresponding true
curve or coefficient. Write
\(\sigma_n^2(c_n)=c_n^\top A_{\Gcal_0}^{-1}\Omega_{\Gcal_0}
A_{\Gcal_0}^{-1}c_n\)
for the population contrast variance.

Let
\(\widehat\lambda_{k,\Gcal}^{(-\ell)}(v,t)
=\widehat\pi_{k,\Gcal}^{(-\ell)}(v\mid t)
\widehat\mu_{k,\Gcal}^{(-\ell)}(v,t)\)
be the cross-fit plug-in for \(\lambda_{k,\Gcal}\), where \(\widehat\pi\)
estimates the conditional density \(\pi_{k,\Gcal}(v\mid t)\) defined above and
\(\widehat\mu\) estimates the conditional mean of \(D_{ik,\Gcal}b_{ik}(v)\), given
\(v_{ik,\Gcal}=v\) and \(T_{ik}=t\).
These
nuisance estimates use observations outside fold \(\ell\).
We omit the fold superscripts below.
In particular, \(\widehat b_{ik,\Gcal}\) is obtained by estimating
the conditional rank density \(b_{ik}(\cdot)\) without fold \(\ell(i)\)
and evaluating that estimate at the spline fitted value for observation
\(i\), computed from a model fitted without the same fold.
Feasible inference uses these plug-ins in the corrected contribution:
\[
\begin{aligned}
\widehat\Delta_{ik,\Gcal}
&=
\int_0^1\widehat\lambda_{k,\Gcal}(v,T_{ik})
\{\one(\widehat R_{ik}\le v)-v\}\dd v,
\\
\widehat A_{\Gcal}
&=
\frac1n\sum_{i=1}^n\sum_{k=1}^K
\widehat b_{ik,\Gcal}
D_{ik,\Gcal}D_{ik,\Gcal}^\top,
\\
\widehat\Phi_{i,\Gcal}
&=
\sum_{k=1}^K
\left[
D_{ik,\Gcal}\psitau\{\widehat R_{ik}
-D_{ik,\Gcal}^\top\widehat\vartheta_{\Gcal}\}
+\widehat\Delta_{ik,\Gcal}
\right].
\end{aligned}
\]
Supplement S.6 gives the plug-in estimators and consistency conditions,
including a threshold on the smallest eigenvalue of the normalized
estimated information matrix to avoid inverting singular or nearly
singular matrices.

When the smallest eigenvalue exceeds the threshold, we compute
\(\widehat q_i=\widehat c_n^\top
\widehat A_{\widehat\Gcal}^{-1}
\widehat\Phi_{i,\widehat\Gcal}\)
as the estimated influence contribution of observation \(i\) to
the chosen contrast. We center these contributions at their average
\(\bar q_n=n^{-1}\sum_{i=1}^n\widehat q_i\). Define
\(\widehat\sigma^2=\frac{1}{n}\sum_{i=1}^n
\{\widehat q_i-\bar q_n\}^2\).
If the smallest eigenvalue is at or below the threshold or \(\widehat\sigma=0\), the reported
interval is \(\mathbb R\). Otherwise, for confidence level
\(1-\alpha\), we draw independent Rademacher multipliers \(e_i\), one for
each observation \(i\), and compute
\[
T^*(c_n)
=
\frac{1}{\sqrt n\widehat\sigma}
\sum_{i=1}^n
e_i\{\widehat q_i-\bar q_n\}.
\]
Let \(z_{1-\alpha}^*\) be the conditional \(1-\alpha\) quantile
of \(|T^*(c_n)|\) given the data. The reported pointwise interval is
\begin{equation}
\label{eq:multiplier-interval}
\left[
\widehat c_n^\top\widehat\vartheta_{\widehat\Gcal}
-n^{-1/2}z_{1-\alpha}^*\widehat\sigma,
\widehat c_n^\top\widehat\vartheta_{\widehat\Gcal}
+n^{-1/2}z_{1-\alpha}^*\widehat\sigma
\right].
\end{equation}
We apply the same multiplier \(e_i\) to the contributions from
all \(K\) responses in observation \(i\), so the bootstrap accounts for their
dependence. We keep the selected partition and nuisance estimates fixed
while drawing the multipliers.

\section{Consistency and Pointwise Inference for Shared Curves}
\label{sec:theory}

We establish recovery of the exact response partition, convergence
rates for the shared curves, and pointwise inference accounting for rank
estimation. We first derive the score and coefficient expansions,
then study curve estimation and partition selection, and finally establish
inference with estimated standard errors. The selection analysis accounts for
both rank-estimation error and the approximation gain from splitting a shared
curve.
We state Conditions C1-C7 in Section~S.1 and the additional inference
conditions, C8-C9, in Section~S.6. These include familiar smoothness,
nondegenerate information, and spline approximation assumptions. They also
specify the accuracy needed to estimate ranks and variances and how to
choose the selection penalty as the spline dimension grows.

\subsection{Generated-Rank Score and Coefficient Expansions}
\label{subsec:main-results}

To obtain convergence rates and confidence intervals, we first need to control
the remainder in the score expansion with estimated ranks.
Theorem~\ref{thm:general-expansion} gives bounds for the full score vector and
for linear contrasts.
Let \(\mathfrak G\) be the fixed finite candidate family containing the true
partition \(\Gcal_0\), and let \(\mathfrak G_+\) consist of \(\Gcal_0\) and
its candidate refinements. A refinement may split a true group but does not
pool responses from different true groups. Define
\(p_n=\max_{\Gcal\in\mathfrak G}d(\Gcal)\). With fixed \(K\) and fixed
covariate dimension, \(p_n\asymp J_n\).

The uniform first-stage rank-error rate is
\(\varepsilon_n=(nh/\log n)^{-1/2}+h+b_y^2\). Conditions C1-C5
specify the sampling and grouped-spline regularity, first-stage smoothing
rates, and local behavior of the quantile-regression criterion. The spline
basis has fixed degree at least two and simple quasi-uniform knots.
Section~S.1 gives the full conditions, including the representer bounds
and their relation to the growing dimension \(p_n\).

We use \(v_{ik,\Gcal}=D_{ik,\Gcal}^\top\vartheta_{\Gcal}^0\),
\(A_{\Gcal}\), and \(\Phi_{i,\Gcal}=\sum_k(S_{ik,\Gcal}+\Delta_{ik,\Gcal})\)
as defined in Subsection~\ref{subsec:corrected-inference}. For a vector \(x\),
write \(\|x\|_A=(x^\top A x)^{1/2}\) and
\(\|x\|_{A^{-1}}=(x^\top A^{-1}x)^{1/2}\). Define the normalized score
remainder
\[
R_{n,\Gcal}^{\rm sc}
=
\frac1{\sqrt n}\sum_{i=1}^n\sum_{k=1}^K
D_{ik,\Gcal}\psitau(\widehat R_{ik}-v_{ik,\Gcal})
-\frac1{\sqrt n}\sum_{i=1}^n\Phi_{i,\Gcal}.
\]

\begin{theorem}[Generated-rank score expansion for grouped splines]
\label{thm:general-expansion}
Assume exact sharing and Conditions C1-C4. Uniformly over
\(\Gcal\in\mathfrak G_+\),
\begin{equation}
\|R_{n,\Gcal}^{\rm sc}\|_{A_{\Gcal}^{-1}}=o_p(\sqrt{p_n}).
\label{eq:general-score}
\end{equation}
For every deterministic sequence \(c_{n,\Gcal}\) with
\(\|c_{n,\Gcal}\|_{A_{\Gcal}^{-1}}=O(1)\), uniformly over the same family,
\begin{equation}
c_{n,\Gcal}^\top A_{\Gcal}^{-1}R_{n,\Gcal}^{\rm sc}=o_p(1).
\label{eq:general-score-direction}
\end{equation}
\end{theorem}

The vector bound is sufficient for obtaining convergence rates for the estimated
curves. For inference on an individual curve value or a coefficient contrast,
the sharper directional bound gives the smaller remainder needed for an
asymptotic distribution. With Condition C5, these score bounds yield the
coefficient expansion in Corollary~\ref{cor:grouped-expansion}. Proofs are given
in Sections~S.2 and~S.3.

\begin{corollary}[Grouped coefficient expansion]
\label{cor:grouped-expansion}
Under exact sharing and Conditions C1-C5, uniformly over
\(\Gcal\in\mathfrak G_+\), every deterministic sequence
\(c_{n,\Gcal}\) satisfying
\(\|c_{n,\Gcal}\|_{A_{\Gcal}^{-1}}=O(1)\) obeys
\begin{equation}
\sqrt n c_{n,\Gcal}^\top
(\widehat\vartheta_{\Gcal}-\vartheta_{\Gcal}^0)
=
c_{n,\Gcal}^\top A_{\Gcal}^{-1}
\frac1{\sqrt n}\sum_{i=1}^n\Phi_{i,\Gcal}+o_p(1),
\label{eq:general-functional}
\end{equation}
and
\begin{equation}
\left\|
\widehat\vartheta_{\Gcal}-\vartheta_{\Gcal}^0
-A_{\Gcal}^{-1}\frac1n\sum_{i=1}^n\Phi_{i,\Gcal}
\right\|_{A_{\Gcal}}
=o_p(\sqrt{p_n/n}).
\label{eq:general-vector}
\end{equation}
\end{corollary}

\begin{remark}[First-stage rates with increasing spline dimension]
\label{rem:sieve-interaction}
Let \(\bar\lambda_n\) measure the size of the score representers, as defined
in Section~S.1.
The rate requirements
\(\sqrt n\bar\lambda_n\varepsilon_n^2\to0\) and
\(\sqrt n\bar\lambda_n(h^2+b_y^2)\to0\), with
\(\bar\lambda_n=O(\sqrt{p_n})\), link the first-stage bandwidths
to spline growth.
The score bounds in \eqref{eq:general-score} and
\eqref{eq:general-score-direction} suffice for \eqref{eq:general-functional}
and \eqref{eq:general-vector} throughout the compatible rate window below.
\end{remark}

\subsection{Curve and Coefficient Estimation}
\label{subsec:grouping-consequences}

Under the true partition, Corollary~\ref{cor:main_curve} gives the convergence
rate of each shared curve and the asymptotic distributions used for inference on
curve values and nuisance coefficients. Condition C6 requires a spline
approximation error of order \(J_n^{-r}\), where \(r\) is the smoothness index
of the true curves, together with the density and spline stability bounds in
Section~S.1.

Let \(\theta_m^0\) and \(\xi_k^s\) be the curve and nuisance blocks of
\(\vartheta_{\Gcal_0}^0\), where superscript \(s\) denotes the spline
pseudo-target. Let \(P_{\xi,k}\) select the nuisance block and let \(e_m(t)\)
map \(\vartheta_{\Gcal_0}\) to \(B(t)^\top\theta_m\). Write
\(\widehat\theta_m\) and \(\widehat\xi_k\) for the corresponding estimated
blocks, set \(\widehat\alpha_m(t)=B(t)^\top\widehat\theta_m\), and let \(w\)
be a fixed integration weight bounded above and away from zero on \([0,1]\).
Define
\(\sigma_{\xi,k,n}^2(c)=c^\top P_{\xi,k}A_{\Gcal_0}^{-1}
\Omega_{\Gcal_0}A_{\Gcal_0}^{-1}P_{\xi,k}^\top c\) and
\(\sigma_{m,n}^2(t)=e_m(t)A_{\Gcal_0}^{-1}
\Omega_{\Gcal_0}A_{\Gcal_0}^{-1}e_m(t)^\top\).

\begin{corollary}[Curve and coefficient consequences]
\label{cor:main_curve}
Under Conditions C1-C6 and exact sharing,
\(
\int_0^1
\{\widehat\alpha_m(t)-\alpha_m^0(t)\}^2w(t)\dd t
=
O_p(J_n/n+J_n^{-2r})
\)
for each group \(m\). For each fixed contrast \(c\in\R^q\), if
\(\sigma_{\xi,k,n}^2(c)\) is bounded above and away from zero,
\[
\frac{\sqrt n c^\top(\widehat\xi_k-\xi_k^s)}
{\sigma_{\xi,k,n}(c)}
\xrightarrow{d}
N(0,1),
\]
where \(\xi_k^s\) is the nuisance block of the spline pseudo-target. If
\(\sqrt nJ_n^{-r}\to0\), the same limit holds with \(\xi_k^0\) replacing
\(\xi_k^s\). For fixed \(t\in(0,1)\),
if \(J_n^{-1}\sigma_{m,n}^2(t)\) is bounded away from zero and infinity, then
\[
\frac{\sqrt n\{\widehat\alpha_m(t)-B(t)^\top\theta_m^0\}}
{\sigma_{m,n}(t)}
\xrightarrow{d} N(0,1).
\]
If the pointwise spline bias is negligible on the \(\sqrt{J_n/n}\) scale, the
same pointwise limit holds when centered at \(\alpha_m^0(t)\).
\end{corollary}

Rank estimation enters the limiting variance through \(\Phi_{i,\Gcal_0}\)
while preserving the curve convergence rate. Section~S.4 proves the corollary.

\subsection{Partition Selection and Refinement Control}
\label{subsec:selection-refinement}

Recovering the true partition requires rejecting both the pooling
of distinct curves and the splitting of a shared curve. These cases follow
the loss-and-penalty comparison for information-criterion consistency
\citep{sin1996information,shao1997asymptotic}, but splitting also raises an
approximation issue. A shared true curve need not lie in the finite spline
space. Its best spline approximation can differ across responses because
their design distributions and density weights in the quantile loss differ.
Separate fits can therefore improve the population loss even with known
ranks. The refinement bound limits this gain to order \(J_n^{-2r}\).
Condition C7 requires the penalty to
dominate this gain and the stochastic loss differences, while remaining too
small to offset the fixed increase in population loss from pooling distinct true curves.
Section~S.1 gives the full condition.

\begin{theorem}[Exact partition and group-number consistency]
\label{thm:main_selection}
Under Conditions C1-C7,
\[
\Pp(\widehat\Gcal=\Gcal_0)\to1,
\qquad
\Pp(\widehat M=M_0)\to1.
\]
If the search is restricted to the true number of groups \(M_0\), the
fixed-\(M_0\) selected partition is also consistent for \(\Gcal_0\).
\end{theorem}

The next corollary compares a refinement with the true partition at their
population spline fits. It bounds the possible reduction in population loss from
splitting a true group, as well as the error from sampling and rank estimation
in this comparison.

\begin{corollary}[Refinement-scale bound]
\label{cor:main_refinement}
Under Conditions C1-C7, uniformly over nontrivial candidate refinements
\(\Gcal\) of \(\Gcal_0\),
\[
Q^R(\Gcal,\vartheta_{\Gcal}^0)
-Q^R(\Gcal_0,\vartheta_{\Gcal_0}^0)
\ge -C J_n^{-2r},
\]
and
\[
\{Q_n^{\widehat R}(\Gcal,\vartheta_{\Gcal}^0)-Q_n^{\widehat R}(\Gcal_0,\vartheta_{\Gcal_0}^0)\}
-\{Q^R(\Gcal,\vartheta_{\Gcal}^0)-Q^R(\Gcal_0,\vartheta_{\Gcal_0}^0)\}
=o_p\left[a_n\{d(\Gcal)-d(\Gcal_0)\}\right].
\]
\end{corollary}

We give the selection proofs in Section~S.5 and verify the rates
in Section~S.8. A larger spline dimension improves curve approximation but
requires tighter control of rank estimation. In the following remark, we
give bandwidth and penalty choices that meet both requirements.

\begin{remark}[A compatible rate window]
\label{sec:rates}
One compatible choice is
\[
h\asymp b_y\asymp n^{-1/3}(\log n)^{1/3},
\qquad
p_n\asymp J_n,
\qquad
\bar\lambda_n=O(\sqrt{J_n}).
\]
With these bandwidth choices, the most restrictive of the stated
upper bounds on \(J_n\) is
\[
J_n=o\{n^{1/3}(\log n)^{-4/3}\}.
\]
When refinements exist, C7 also requires \(n a_n\to\infty\),
\(a_nJ_n\to0\), and \(a_nJ_n^{2r+1}\to\infty\), hence
\(J_n\gg a_n^{-1/(2r+1)}\). For \(a_n\asymp\log n/n\), the resulting lower
bound is compatible with the displayed upper bound when \(r>1\). This
threshold is specific to that penalty scale. For inference on a fixed nuisance contrast centered at its true value,
\(\sqrt nJ_n^{-r}\to0\) is a sufficient bias condition. This condition is
compatible with the displayed upper bound when \(r>3/2\). Pointwise inference
centered at the true curve also requires negligible spline bias. Section~S.8 verifies these rate restrictions.
\end{remark}

\subsection{Pointwise Inference under Consistent Selection}
\label{subsec:inference-after-selection}

Having established consistent group selection, we now consider confidence
intervals computed from the selected fit. Theorem~\ref{thm:main_post} shows that
estimated standard errors and multiplier critical values give valid pointwise
inference. The contrasts and intervals are
defined in Subsections~\ref{subsec:corrected-inference}
and~\ref{subsec:feasible-inference}. Conditions C8-C9, stated in
Section~S.6, specify the additional requirements for inference. Condition C8
requires relative consistency of the estimated information matrix and score
representer. Condition C9 requires a nondegenerate normalized contrast and
independent multipliers, one for each observation \(i\), with finite
\(2+\nu\) moment, for some
\(\nu>0\).

\begin{theorem}[Pointwise inference after consistent partition selection]
\label{thm:main_post}
Under Conditions C1-C9, suppose
\(
\norm{c_n/\sigma_n(c_n)}_{A_{\Gcal_0}^{-1}}=O(1)
\)
with \(\sigma_n(c_n)>0\). Then
\[
\frac{\sqrt n\{\widehat c_n^\top\widehat\vartheta_{\widehat\Gcal}
-c_n^\top\vartheta_{\Gcal_0}^0\}}{\sigma_n(c_n)}
=
\frac{c_n^\top A_{\Gcal_0}^{-1}}{\sigma_n(c_n)}
\frac{1}{\sqrt n}\sum_{i=1}^n\Phi_{i,\Gcal_0}
+o_p(1).
\]
The \(\widehat\sigma\)-studentized statistic is asymptotically standard normal.
With \(z_{1-\alpha}^*\) the conditional \(1-\alpha\) quantile of
\(|T^*(c_n)|\), the interval
\[
\widehat c_n^\top\widehat\vartheta_{\widehat\Gcal}
\mathbin{\pm}
n^{-1/2}z_{1-\alpha}^*\widehat\sigma
\]
has asymptotic pointwise coverage \(1-\alpha\).
\end{theorem}

Theorem~\ref{thm:main_post} justifies pointwise confidence intervals after
estimating both the group number and membership. Their standard errors include
the contribution from rank estimation and dependence among the
\(K\) responses in each observation. For curve contrasts, coverage concerns the
spline pseudo-curve
unless pointwise spline bias is negligible. Validity is pointwise in the
exact-sharing distribution, rather than uniform near a
selection boundary or simultaneous across selected submodels
\citep{leeb2005model,berk2013valid}. Section~S.6 gives the proof.

\section{Simulation Studies of Grouping and Inference}
\label{sec:simulation_generated_rank}

We examine whether joint estimation improves recovery of response groups and
their shared coefficient curves when both the group number and membership are
unknown. A separate experiment evaluates pointwise inference after selecting
the partition. The study spans three grouping settings, three signal levels,
and four sample sizes.

\subsection{Design, Comparators, and Performance Measures}
\label{subsec:simulation_dgm}

Ten responses form three true groups of sizes \((2,2,6)\) under a grouped
conditional-median model. We vary true-curve separation through three settings
and use signal multiplier \(\omega\in\{0.75,1,1.5\}\). Setting 2 has the
weakest separation, whereas Setting 3 has the clearest separation.
Within this section, \(\alpha_m\) denotes the analytic true curve for group
\(m\).

The design combines heterogeneous observed responses with within-subject
dependence while preserving the grouped conditional-median model on the rank
scale. All procedures use the same five-fold cross-fitted conditional ranks
and a cubic B-spline basis with \(J_n=4\) at every sample size. Each
combination of sample size, setting, and signal level uses 500 Monte Carlo replications. Supplement
S.9 gives the data-generating distributions,
curve specifications, and estimation details.

We compare the proposed method, which estimates group membership and shared
curves jointly, with Rank-QR plus K-means. The latter first fits a separate
quantile regression to each response's estimated conditional ranks. For each
candidate group number, it clusters the fitted spline coefficient vectors and
refits the grouped model under the resulting partition. Both methods select the
group number from \(M\in\{1,\ldots,10\}\) using the same form of the information
criterion and the same penalty factor. Supplement~S.9 reports comparisons with
\(M=M_0=3\) fixed, along with curve errors for Marginal Rank-QR and the
true-partition reference. These two reference procedures fit responses
separately or pool them according to their known group membership, respectively,
and both use estimated conditional ranks.

We measure group recovery by Exact, the proportion of replications in which
the estimated partition equals \(\mathcal G_0\) up to relabeling. It requires
correct recovery of both the group number and every response's membership.
For curve estimation, let \(\widehat\alpha_k\) denote the fitted rank-scale
curve for the \(k\)th response. We report \(10^3\) times the Monte Carlo mean of the
integrated squared error averaged over responses, defined for one replication by
\[
\operatorname{ISE}
=
\frac{1}{K}\sum_{k=1}^K\int_0^1
\{\widehat\alpha_k(t)-\alpha_{g_0(k)}(t)\}^2\mathrm{d}t.
\]
\subsection{Partition Recovery and Rank-Scale Curve Estimation}
\label{subsec:simulation_grouping_results}

The proposed method improves average partition recovery and curve estimation
over Rank-QR plus K-means at each sample size considered. Across the 36
combinations of sample size, setting, and signal level in Table~\ref{tab:simulation_summary}, it
has higher exact recovery in 33 combinations and lower mean curve ISE in 33
combinations. The two small reversals in exact recovery occur at the largest
sample size and strongest signal. The proposed method also has higher average
rates of selecting the correct group number and assigning responses to their
true groups (Supplement~S.9). The average group-number and membership recovery
of Rank-QR plus K-means remains lower across the penalty constants examined
there.

When the true group number is supplied, the proposed method retains its
advantage in partition recovery over Rank-QR plus K-means. Its curve-estimation
error is lower than that of Marginal Rank-QR in all 36 combinations, and the
gap in average error between the proposed method and the true-partition
reference decreases with sample size (Supplement~S.9).

Recovery is more difficult when the true group curves are close. At
\(n=50{,}000\), the proposed method recovers the entire partition in more than
90\% of replications in Settings 1 and 3. In Setting 2, recovery at this sample
size is 0.040, 0.270, and 0.920 for \(\omega=0.75,1,1.5\), respectively. A large
sample may therefore still be insufficient to distinguish a close pair of
curves. Figure~\ref{fig:simulation_recovery_separation} shows the same pattern
across sample sizes using
\(d_{\min}=\min_{a<b}\{\int_0^1
[\alpha_a(t)-\alpha_b(t)]^2\mathrm{d}t\}^{1/2}\),
the minimum pairwise \(L_2\) separation among the true group curves. Exact
recovery generally increases with both curve separation and sample size.

Selecting the group number remains a source of error even when the curves are
well separated. At \(n=50{,}000\), fixing \(M=M_0\) gives exact recovery of
1.000 in Settings 1 and 3 and in Setting 2 with \(\omega=1.5\). When \(M\) is
selected, recovery in these seven designs ranges from 0.912 to 0.962. All
remaining recovery failures in these designs occur when the criterion selects
more than three groups.

Curve-estimation errors decrease with sample size for both methods.
At \(n=50{,}000\), the two methods have similar errors in Settings 1 and 3
at the strongest signal.

\begin{table}[!p]
\centering
\caption{Partition recovery and curve estimation when both methods select the
group number. Each entry uses 500 replications. K-means denotes Rank-QR plus
K-means.}
\label{tab:simulation_summary}
\begingroup
\renewcommand{\baselinestretch}{1}
\footnotesize
\setlength{\tabcolsep}{2pt}
\begin{tabular*}{\textwidth}{@{\extracolsep{\fill}}rr*{8}{r}@{}}
\toprule
\multicolumn{10}{c}{Panel A: Exact partition recovery} \\
\addlinespace[2pt]
Setting & Signal & \multicolumn{2}{c}{\(n=2{,}000\)} & \multicolumn{2}{c}{\(n=5{,}000\)} & \multicolumn{2}{c}{\(n=10{,}000\)} & \multicolumn{2}{c}{\(n=50{,}000\)} \\
\cmidrule(lr){3-4}\cmidrule(lr){5-6}\cmidrule(lr){7-8}\cmidrule(lr){9-10}
 & & Proposed & K-means & Proposed & K-means & Proposed & K-means & Proposed & K-means \\
\midrule
1 & 0.75 & 0.022 & 0.000 & 0.252 & 0.000 & 0.586 & 0.012 & 0.912 & 0.252 \\
 & 1.00 & 0.112 & 0.000 & 0.498 & 0.004 & 0.742 & 0.034 & 0.914 & 0.636 \\
 & 1.50 & 0.410 & 0.002 & 0.738 & 0.028 & 0.798 & 0.152 & 0.918 & 0.942 \\
\addlinespace[3pt]
2 & 0.75 & 0.002 & 0.000 & 0.000 & 0.000 & 0.008 & 0.002 & 0.040 & 0.006 \\
 & 1.00 & 0.004 & 0.000 & 0.012 & 0.000 & 0.016 & 0.002 & 0.270 & 0.010 \\
 & 1.50 & 0.034 & 0.000 & 0.128 & 0.002 & 0.424 & 0.014 & 0.920 & 0.136 \\
\addlinespace[3pt]
3 & 0.75 & 0.056 & 0.000 & 0.340 & 0.004 & 0.638 & 0.034 & 0.918 & 0.288 \\
 & 1.00 & 0.258 & 0.000 & 0.724 & 0.026 & 0.818 & 0.090 & 0.932 & 0.786 \\
 & 1.50 & 0.654 & 0.024 & 0.790 & 0.190 & 0.840 & 0.462 & 0.962 & 0.980 \\
\bottomrule
\end{tabular*}
\par\smallskip
\begin{tabular*}{\textwidth}{@{\extracolsep{\fill}}rr*{8}{r}@{}}
\toprule
\multicolumn{10}{c}{Panel B: Curve-estimation error, \(10^3\times\) mean ISE} \\
\addlinespace[2pt]
Setting & Signal & \multicolumn{2}{c}{\(n=2{,}000\)} & \multicolumn{2}{c}{\(n=5{,}000\)} & \multicolumn{2}{c}{\(n=10{,}000\)} & \multicolumn{2}{c}{\(n=50{,}000\)} \\
\cmidrule(lr){3-4}\cmidrule(lr){5-6}\cmidrule(lr){7-8}\cmidrule(lr){9-10}
 & & Proposed & K-means & Proposed & K-means & Proposed & K-means & Proposed & K-means \\
\midrule
1 & 0.75 & 3.900 & 5.066 & 1.465 & 2.346 & 0.608 & 1.107 & 0.088 & 0.132 \\
 & 1.00 & 3.929 & 5.672 & 1.298 & 2.329 & 0.517 & 1.042 & 0.087 & 0.104 \\
 & 1.50 & 3.365 & 5.909 & 1.028 & 2.018 & 0.472 & 0.717 & 0.086 & 0.083 \\
\addlinespace[3pt]
2 & 0.75 & 3.340 & 2.808 & 1.378 & 1.573 & 0.588 & 1.031 & 0.105 & 0.131 \\
 & 1.00 & 3.512 & 3.654 & 1.266 & 2.069 & 0.557 & 0.942 & 0.132 & 0.151 \\
 & 1.50 & 3.433 & 5.228 & 1.359 & 1.932 & 0.636 & 0.954 & 0.086 & 0.160 \\
\addlinespace[3pt]
3 & 0.75 & 3.167 & 5.236 & 1.251 & 1.825 & 0.568 & 0.897 & 0.086 & 0.131 \\
 & 1.00 & 3.273 & 5.051 & 1.046 & 1.952 & 0.467 & 0.857 & 0.084 & 0.094 \\
 & 1.50 & 2.580 & 4.804 & 0.942 & 1.390 & 0.448 & 0.553 & 0.080 & 0.078 \\
\bottomrule
\end{tabular*}
\endgroup
\medskip
\begin{minipage}{\textwidth}
\captionsetup{type=figure}
\centering
\includegraphics[width=0.82\textwidth]{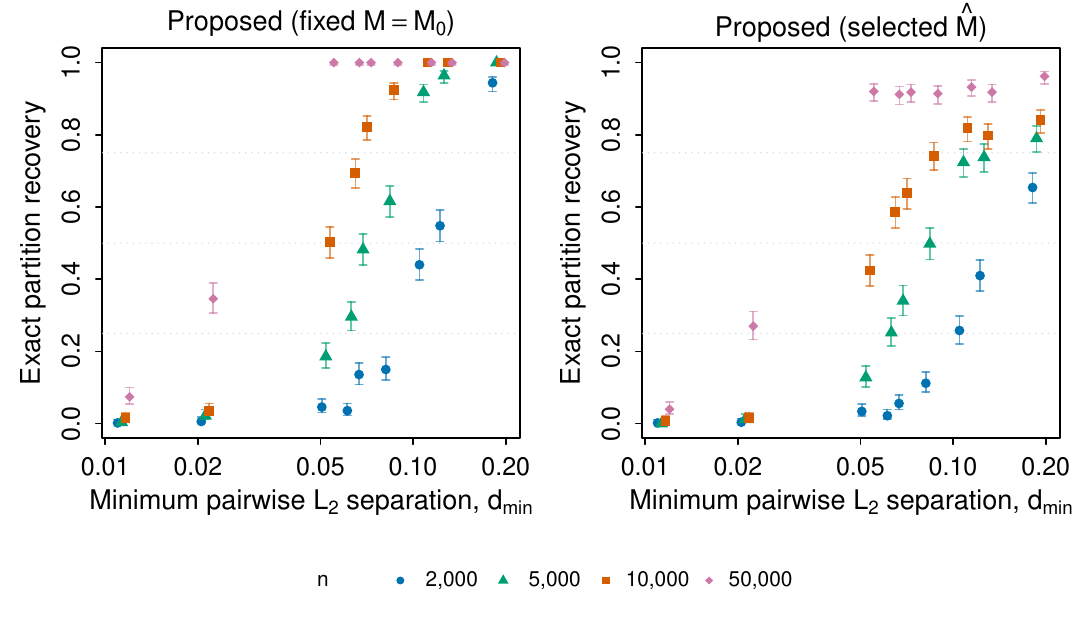}
\caption{Exact partition recovery versus minimum true-curve separation for the
proposed estimator with \(M=M_0\) fixed (left) or \(M\) selected (right). Points
are
500-replication recovery proportions, bars are 95\% Wilson intervals, and the
sample sizes are displaced slightly for visibility.}
\label{fig:simulation_recovery_separation}
\end{minipage}
\end{table}

\subsection{Pointwise Interval Calibration after Partition Selection}
\label{subsec:simulation_inference}

We next assess pointwise intervals after re-estimating the group number and
membership in every replication. Table~\ref{tab:simulation_inference}
reports results for the spline pseudo-curve
\(B(t)^\top\theta_1^0\) for the first response variable at \(t=0.50\) in all three settings.
At each signal level, its true coefficient curve is the same across
settings, while the other group curves and the difficulty of partition
recovery vary.
Bias is the Monte Carlo mean of
\(\widehat\alpha_1(t)-B(t)^\top\theta_1^0\), SE averages the estimated
pointwise standard errors, and Coverage is the proportion of 95\% pointwise
intervals containing \(B(t)^\top\theta_1^0\). Supplement S.9 gives the
corresponding results at \(t=0.25\) and \(0.75\) and the separate
known-membership experiment.

Coverage generally improves with sample size. At \(n=2{,}000\), signal
\(\omega=0.75\), and \(t=0.50\), coverage is 0.790, 0.658, and 0.832 in Settings
1, 2, and 3, respectively. Mean estimated standard errors are also smaller than
the empirical standard deviations across replications: their ratios range from
0.629 to 0.923 across the 18 cases at \(t=0.50\) with \(n=2{,}000\) or
\(5{,}000\). Supplement~S.9 reports these empirical standard deviations and the
corresponding ratios.
At \(n=50{,}000\), coverage across all settings, signal levels, and evaluation
points ranges from 0.950 to 0.964, with absolute bias no larger than 0.00082.
The estimated standard errors also closely match the empirical standard
deviations at \(t=0.50\) (Supplement~S.9). At this sample size, the intervals
for the first response variable have near-nominal coverage even in Setting 2 at the weaker
signals, where the full partition remains difficult to recover. These intervals
concern the spline pseudo-curve for the first response variable, whereas exact partition
recovery requires the correct grouping of all ten responses. In the separate
experiment with known membership, bias is small and the estimated standard
errors closely track the empirical standard deviations (Supplement~S.9).

\begin{table}[!htbp]
\centering
\caption{Pointwise estimation and interval calibration after partition selection
at \(t=0.50\) for the spline pseudo-curve for the first response variable. SE is the mean estimated
standard error, and Cov. is empirical coverage of 95\% pointwise intervals.}
\label{tab:simulation_inference}
\renewcommand{\baselinestretch}{1}
\footnotesize
\setlength{\tabcolsep}{2pt}
\begin{tabular*}{\textwidth}{@{\extracolsep{\fill}}r*{12}{r}@{}}
\toprule
Signal & \multicolumn{3}{c}{\(n=2{,}000\)} & \multicolumn{3}{c}{\(n=5{,}000\)} & \multicolumn{3}{c}{\(n=10{,}000\)} & \multicolumn{3}{c}{\(n=50{,}000\)} \\
\cmidrule(lr){2-4}\cmidrule(lr){5-7}\cmidrule(lr){8-10}\cmidrule(lr){11-13}
 & Bias & SE & Cov. & Bias & SE & Cov. & Bias & SE & Cov. & Bias & SE & Cov. \\
\midrule
\multicolumn{13}{l}{Setting 1} \\
0.75 & 0.0244 & 0.0323 & 0.790 & 0.0127 & 0.0226 & 0.760 & 0.0044 & 0.0179 & 0.854 & -0.0001 & 0.0084 & 0.950 \\
1.00 & 0.0251 & 0.0341 & 0.744 & 0.0073 & 0.0247 & 0.842 & 0.0012 & 0.0185 & 0.928 & -0.0001 & 0.0083 & 0.954 \\
1.50 & 0.0110 & 0.0380 & 0.830 & 0.0012 & 0.0252 & 0.918 & 0.0010 & 0.0179 & 0.952 & -0.0002 & 0.0081 & 0.952 \\
\midrule
\multicolumn{13}{l}{Setting 2} \\
0.75 & 0.0291 & 0.0325 & 0.658 & 0.0136 & 0.0232 & 0.760 & 0.0035 & 0.0182 & 0.882 & -0.0001 & 0.0084 & 0.950 \\
1.00 & 0.0259 & 0.0353 & 0.692 & 0.0064 & 0.0253 & 0.868 & 0.0008 & 0.0186 & 0.940 & -0.0001 & 0.0083 & 0.954 \\
1.50 & 0.0095 & 0.0386 & 0.846 & 0.0010 & 0.0253 & 0.922 & 0.0010 & 0.0179 & 0.952 & -0.0002 & 0.0081 & 0.952 \\
\midrule
\multicolumn{13}{l}{Setting 3} \\
0.75 & 0.0231 & 0.0355 & 0.832 & 0.0125 & 0.0241 & 0.790 & 0.0043 & 0.0183 & 0.850 & -0.0001 & 0.0084 & 0.950 \\
1.00 & 0.0210 & 0.0380 & 0.778 & 0.0025 & 0.0261 & 0.916 & 0.0007 & 0.0186 & 0.950 & -0.0001 & 0.0083 & 0.954 \\
1.50 & 0.0038 & 0.0400 & 0.906 & 0.0009 & 0.0252 & 0.922 & 0.0010 & 0.0179 & 0.952 & -0.0002 & 0.0081 & 0.952 \\
\bottomrule
\end{tabular*}
\end{table}

\section{Application to UK Biobank Data}
\label{sec:ukb-analysis}

Genetic susceptibility to type 2 diabetes (T2D) is associated with several
aspects
of metabolic health, including adiposity, glucose regulation, lipids, and
blood pressure \citep{udler2018softclustering,dicorpo2022t2dprs}.
Age-varying associations are of particular interest in women during midlife,
when the menopause transition is accompanied by changes in body composition
\citep{greendale2019menopause}. We use UK Biobank data to ask how metabolic
traits group according to their associations with inherited susceptibility
to type 2 diabetes at lower, middle, and upper conditional quantiles. This
comparison can reveal similarities in the tails that a study of central
associations alone would miss, while the coefficient curves describe how
each association varies with age.

The analysis included 127{,}040 women aged 40 to
70 with complete required measurements from UK Biobank
\citep{sudlow2015ukbiobank}. We excluded participants flagged as having
a genetic relative and those with unknown relatedness status. We did not
restrict the sample by menopausal status. The predictor was the UK Biobank
Standard
type 2 diabetes polygenic risk score
(PRS). This score summarizes inherited susceptibility using genetic
associations estimated in external genome-wide association studies
\citep{thompson2024ukbprs,ukbiobank26285}. We examined body mass index (BMI)
and waist-to-hip ratio (WHR) as measures of body size and fat distribution.
The remaining responses were glucose (GLU), hemoglobin A1c
(HbA1c), systolic blood pressure (SBP), and four lipid measures:
triglycerides (TRIG), high-density lipoprotein cholesterol (HDL),
apolipoprotein B (ApoB), and low-density lipoprotein cholesterol (LDL).

We expressed each observed value as a percentile rank among
women of the same age, placing the nine responses on a common scale. The
age-specific reference distribution for each response averages over PRS and
genetic principal components.
We then fitted
quantile regressions of these ranks on the standardized PRS, adjusting for
the first ten standardized genetic principal components, without an
ancestry-based exclusion. The levels \(\tau=0.1,0.5,0.9\) describe the lower,
middle, and upper parts of each rank distribution conditional on PRS, age,
and these adjustment variables. They describe conditional
distributions and do not divide women into fixed groups with low, middle, or
high response values. HDL was negated before ranking, so a positive
coefficient indicates an association with lower HDL. The lower and upper
quantiles for reversed HDL correspond to the upper and lower quantiles for
HDL, respectively.

We estimated the reference distributions by five-fold cross-fitting and
selected the group number and partition separately at each quantile. The
curves were evaluated over ages 43 to 67 with 95\% pointwise intervals that
include the rank-estimation correction and keep the selected partition fixed.
Supplement S.10 gives the computational details. The age profiles compare
women of different ages and do not estimate within-person changes through
menopause.

\begin{figure}[!htbp]
\centering
\includegraphics[width=\textwidth]{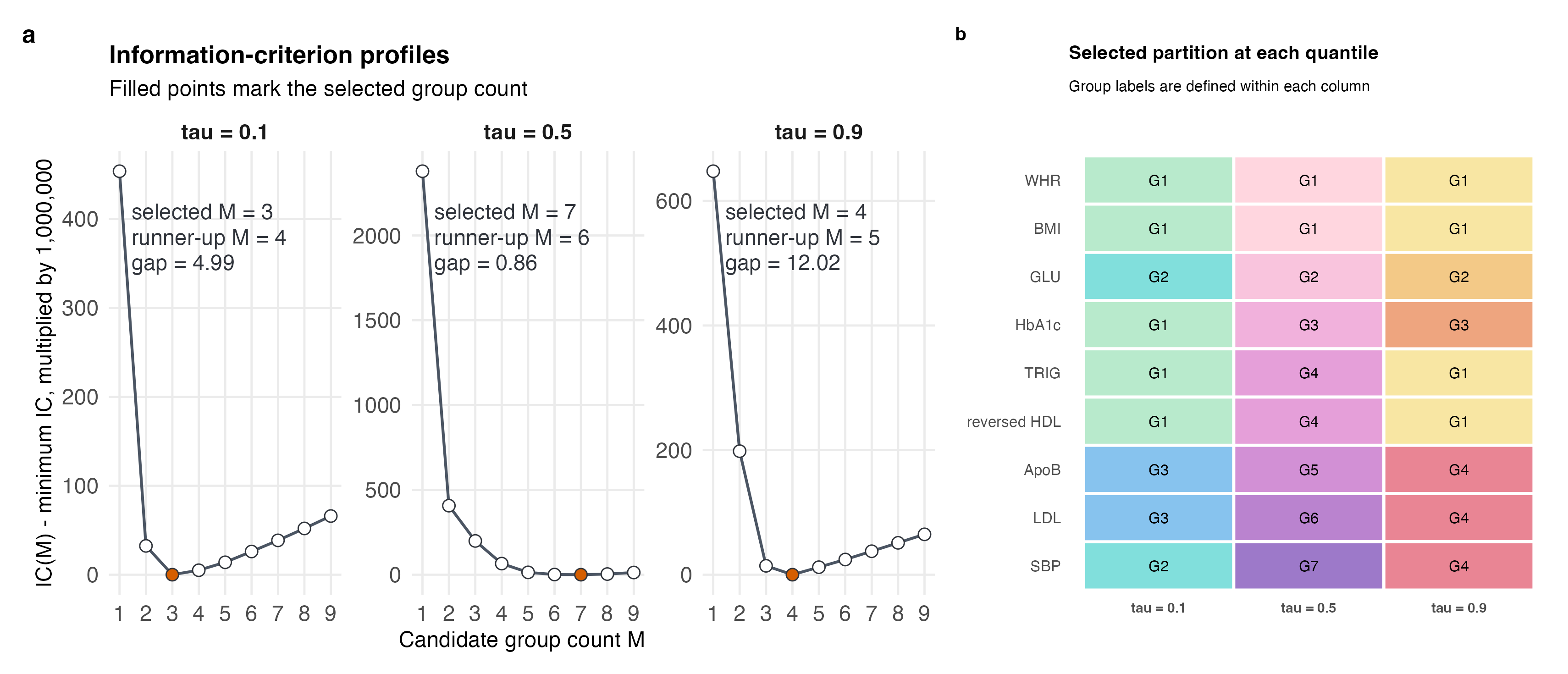}
\caption{Grouping responses by their age-varying associations
with the type 2 diabetes PRS. Panel (a) shows the information-criterion profiles, and
panel (b) shows the selected response groups. Both panels report results at
\(\tau=0.1,0.5,0.9\).}
\label{fig:ukb-selection}
\end{figure}

The information criterion selected seven groups at the median, compared with
three at the lower quantile and four at the upper quantile
(Figure~\ref{fig:ukb-selection}), from the full candidate range
\(M=1,\ldots,9\). At the median, BMI and WHR form one pair and triglycerides
and reversed HDL another; each remaining trait has its own curve. In the
tails, more traits share a curve, with different groups at the lower and
upper quantiles.

\begin{figure}[H]
\centering
\includegraphics[width=\textwidth]{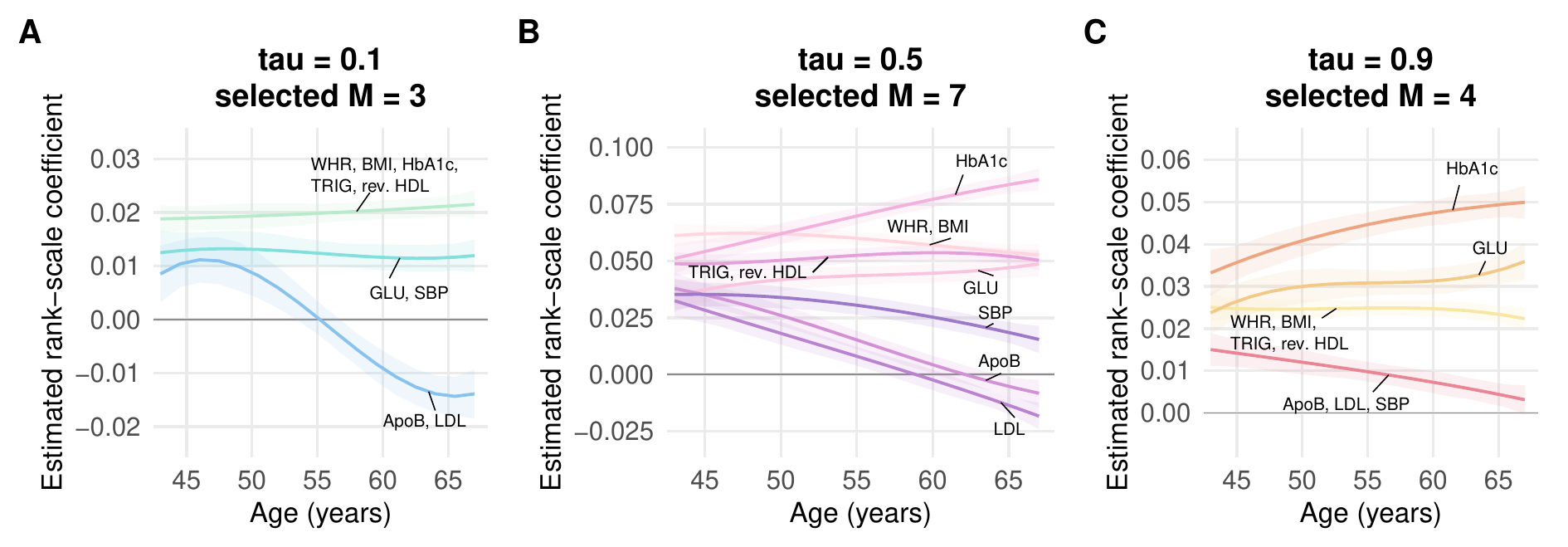}
\caption{Selected PRS coefficient curves on the age-specific percentile scale
at \(\tau=0.1,0.5,0.9\). Shaded regions show 95\% pointwise confidence
intervals.}
\label{fig:ukb-curves}
\end{figure}

At each of the three quantiles, BMI and WHR shared a fitted PRS curve, as did
triglycerides and reversed HDL. These two pairs formed a single group in both
tails but remained separate at the median. Higher genetic susceptibility was
associated with higher BMI, WHR, and triglycerides and lower HDL
(Figure~\ref{fig:ukb-curves}). For the lipid pair, this corresponds to the
combination of higher triglycerides and lower HDL associated with insulin
resistance \citep{oliveri2024tghdl,deforest2024tghdl}. HbA1c, which reflects
glycemic and nonglycemic genetic influences \citep{soranzo2010hba1c}, joined
these four traits only at the lower quantile. It had its own curve at the median
and upper quantile, and its fitted PRS curve was distinct from the glucose curve
at all three quantiles.
Systolic blood pressure grouped with glucose at the lower quantile and with ApoB
and LDL at the upper quantile, but had its own curve at the median. ApoB and LDL
measure different aspects of atherogenic lipid burden \citep{sniderman2024apob}
and shared a fitted PRS curve in both tails despite having separate median
curves.

The fitted associations also varied with age. At the median and upper quantile,
a fixed difference in PRS corresponded to a larger difference in the conditional
quantiles of rank-standardized glucose and HbA1c at older ages, holding the
adjustment variables fixed. By contrast, the positive upper-quantile association
shared by ApoB, LDL, and systolic blood pressure weakened toward zero at older
ages. At the lower quantile, the shared ApoB-LDL coefficient was positive at
younger ages and negative at older ages. Its 95\% pointwise interval was
entirely above zero at age 43 and entirely below zero at age 67, with intervals
including zero at intermediate ages. Most other selected curves had pointwise
intervals above zero at every evaluated age. The separate median ApoB and LDL
curves also had intervals including zero at some evaluated ages, and the
upper-quantile ApoB-LDL-SBP interval included zero at age 67.

Several traits with distinct median associations shared fitted PRS coefficient
curves in the tails, with different groupings at the lower and upper quantiles.
A median-only analysis or a common partition across quantiles would miss these
patterns. The shared curves describe similarities in PRS associations on the
age-specific percentile scale.

\section{Discussion}
\label{sec:discussion}

We have developed a method that jointly identifies responses with a common
quantile association curve and estimates that curve. Responses are compared on
estimated conditional percentile scales, with groups allowed to differ across
quantile levels. The theoretical analysis establishes consistent group recovery
and feasible pointwise inference, accounting for both percentile estimation
and spline approximation.

Reusing subset fits reduces repeated estimation, but the number of subsets
still grows rapidly with the number of responses, making an
exhaustive search impractical for very large phenotype collections. Restricting
the search can change the selected partition and the responses pooled for
estimation, with consequences for both the fitted curves and their inference.
Extending the method to large phenotype collections therefore requires
computationally efficient searches whose statistical properties can also be
established.

Another potential future direction is to extend this approach to an online
estimation and inference framework for studies that continuously recruit new
participants. New participants contribute to the reference distributions and may
change both the percentile values assigned to earlier observations and the
selected response groups. An online extension would need to reuse previous fits
while accommodating these changes and providing valid inference for the updated
curves.

\section*{Acknowledgments}
This research has been conducted using the UK Biobank Resource under
application number 207159.

\phantomsection
\label{main:references-start}
\bibliographystyle{apalike}
\bibliography{references}

\end{document}